\documentclass[pra,showpacs,superscriptaddress,twocolumn]{revtex4-2}

\usepackage{graphicx}
\usepackage{amsmath}
 
\graphicspath{{graphics/}}  
\usepackage{color}
\usepackage{cancel}
\usepackage{ulem}
\usepackage{comment} 
\usepackage{bm} 
\usepackage{braket}
\usepackage{amsmath}
\usepackage{amssymb}

\usepackage{float}

\begin{document}
\title{Coexistence of Magnon-Induced Optical Vortex and Gaussian Beam Scattering Assisted by Rotational Umklapp Process}

\author{Ryusuke~Hisatomi}
\email{hisatomi.ryusuke.2a@kyoto-u.ac.jp}
\affiliation{Institute for Chemical Research (ICR), Kyoto University, Gokasho, Uji, Kyoto 611-0011, Japan}
\affiliation{Center for Spintronics Research Network (CSRN), Kyoto University, Gokasho, Uji, Kyoto 611-0011, Japan}
\author{Kotaro~Taga}
\affiliation{SANKEN, The University of Osaka, Mihogaoka, Ibaraki, Osaka 567-0047, Japan}
\author{Hisakazu~Matsuki}
\affiliation{Institute for Chemical Research (ICR), Kyoto University, Gokasho, Uji, Kyoto 611-0011, Japan}
\affiliation{Center for Spintronics Research Network (CSRN), Kyoto University, Gokasho, Uji, Kyoto 611-0011, Japan}
\author{Shutaro~Karube}
\affiliation{Institute for Chemical Research (ICR), Kyoto University, Gokasho, Uji, Kyoto 611-0011, Japan}
\affiliation{Center for Spintronics Research Network (CSRN), Kyoto University, Gokasho, Uji, Kyoto 611-0011, Japan}
\author{Yoichi~Shiota}
\affiliation{Institute for Chemical Research (ICR), Kyoto University, Gokasho, Uji, Kyoto 611-0011, Japan}
\affiliation{Center for Spintronics Research Network (CSRN), Kyoto University, Gokasho, Uji, Kyoto 611-0011, Japan}
\author{Teruo~Ono}
\affiliation{Institute for Chemical Research (ICR), Kyoto University, Gokasho, Uji, Kyoto 611-0011, Japan}
\affiliation{Center for Spintronics Research Network (CSRN), Kyoto University, Gokasho, Uji, Kyoto 611-0011, Japan}
\affiliation{International Center for Synchrotron Radiation Innovation Smart, Tohoku University, Sendai, Miyagi 980-8577, Japan}

\date{\today}

\begin{abstract}
The exploitation of crystal-lattice symmetries to engineer angular momentum transfer between structured light and magnons marks a novel frontier for optomagnonic research. 
In Brillouin light scattering, when focused light propagates parallel to an external magnetic field and interacts with ferromagnetic uniform magnons, only optical-vortex scattering is expected to be permitted. 
Due to the combined effects of magneto-optical coupling and optical spin-orbit interaction, the transfer of magnon spin angular momentum to photon orbital angular momentum allows for this distinctive scattering phenomenon. 
Here, we experimentally demonstrate that, for a specific ferromagnetic crystal orientation, Gaussian-beam scattering coexists with the optical-vortex scattering, contrary to conventional expectations based on angular momentum conservation between magnons and photons. We show that the crystal lattice, via the rotational Umklapp process, provides the missing angular momentum required for the Gaussian-beam scattering. Furthermore, we predict that as the degree of light focusing increases, the relative efficiencies of the Gaussian-beam and optical-vortex scattering processes reverse.
\end{abstract}


\maketitle

\begin{figure}[t]
\begin{center}
\includegraphics[width=8.0cm,angle=0]{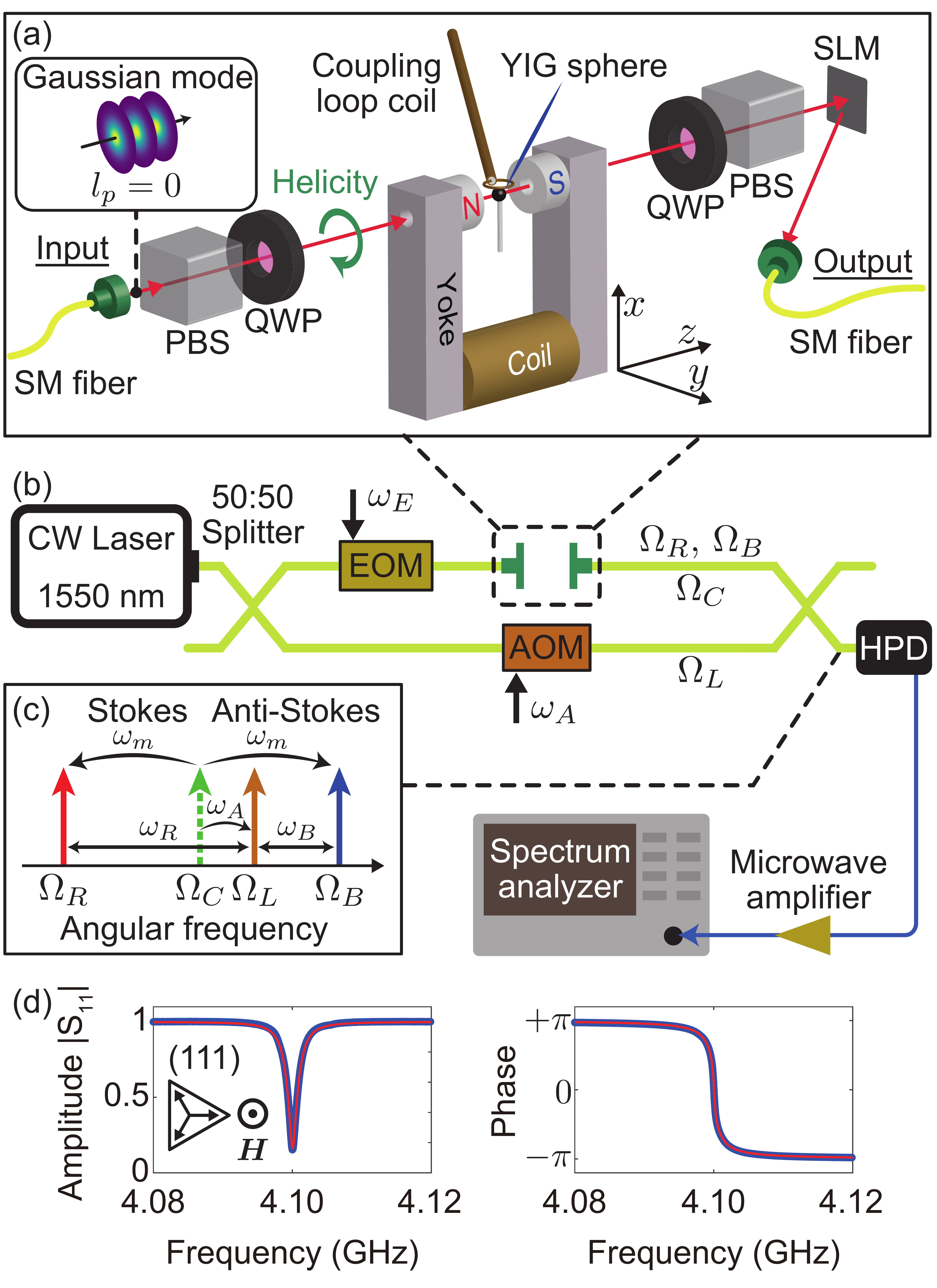}
\caption{
Experimental setup and characterization of the uniform magnon mode. (a)~Optical configuration for free-space light propagation. A spherical YIG single crystal ($0.5\text{-}\text{mm}$ diameter) is placed in a magnetic circuit ($\bm{H} \parallel \langle111\rangle \parallel z$, replacing the $\langle 100\rangle$ axis~\cite{RA2025}). A single-loop coil excites the uniform magnon (Kittel) mode. Quarter-wave plates (QWPs) and polarizing beam splitters (PBSs) resolve the light helicities. An incident Gaussian beam focused to $\sim 100\,\mu\text{m}$ is injected into the YIG sphere; coaxially scattered light is reflected by a spatial light modulator (SLM) and coupled into a single-mode (SM) fiber to identify the optical orbital angular momentum (OAM). (b)~Optical heterodyne detection system. Light from a continuous-wave (CW) laser is split into calibration (via EOM) and local oscillator (LO, via AOM) paths, recombined to generate an optical beat, and analyzed via a high-speed photodetector (HPD) and a spectrum analyzer. (c)~Frequency schematics of the carrier ($\Omega_C$), scattered sidebands ($\Omega_R, \Omega_B$), and beat signals ($\omega_R, \omega_B$). (d)~Microwave reflection spectrum ($|\text{S}_{11}|$) of the uniform magnon mode, showing measured data (blue) and fits (red)~\cite{RA2016}.
} 
\label{fig:setup111}
\end{center}
\end{figure}

In physical processes exhibiting continuous rotational symmetry, angular momentum serves as a good quantum number~\cite{JJ1994}.
For example, the angular momentum of a boson is quantized in integer multiples of the reduced planck constant,~$\hbar$.
Under these conditions, angular momentum can be transferred between distinct systems while conserving the total angular momentum of the entire system.
This angular momentum transfer is fundamental to the advancement of optomagnonics and the emerging field of cavity optomagnonics~\cite{TY2012,RA2016,SH2016,ZN2016,AR2016,JA2016,DY2020}.
Indeed, experimental demonstrations have established the bidirectional transfer of angular momentum between photons and magnons~\cite{AR2016,RA2019,AA2018}.

Specifically, it has been shown that when a focused light beam propagates parallel to an external magnetic field and interacts with a uniform ferromagnetic magnon, the scattering process involves a change in the photon’s orbital angular momentum (OAM)~\cite{LM1992}, converting the incident Gaussian beam into a scattered optical vortex beam~\cite{RA2025}.
Because the photon spin states (helicity) are restricted to $\pm1\hbar$, scattering by a single uniform magnon (spin $1\hbar$~\cite{TH1940,MD1986}) dictates a corresponding change in the photon’s OAM, irrespective of whether the incident helicity is conserved or inverted.
Importantly, this optical-vortex scattering is governed by the interplay between magneto-optical coupling—such as the Faraday and Cotton-Mouton effects—and the optical spin-orbit interaction, whose efficiency fundamentally scales with the degree of light focusing.

However, in crystalline solids, the continuous rotational symmetry of free space is broken down into discrete rotational symmetries, giving rise to the rotational Umklapp process~\cite{RA2019,HN1968,N1980,JE2002,TN2011,KT2014}. 
Within optomagnonics and cavity optomagnonics, the potential synergy among magneto-optical coupling, optical spin-orbit interactions, and such crystal-symmetry-driven Umklapp processes has so far received no attention.
When focused light interacts with magnons within a crystalline lattice, these combined effects are expected to manifest and fundamentally modify the conventional selection rules.

In this paper, we present an experimental investigation of Brillouin light scattering (BLS) arising from the interaction between focused light and magnons in a spherical ferromagnetic crystal. 
To preserve the required rotational symmetries, the experiment was performed under the Faraday geometry, in which both the incident light propagation vector and the external magnetic field are aligned parallel to a specific crystal axis. 
Utilizing an optical heterodyne detection scheme sensitive to both helicity and optical vortices~\cite{RA2016,RA2019,RA2025}, we demonstrate the coexistence of two distinct BLS processes: Gaussian-beam scattering, which involves a change in helicity without altering the photon's OAM, and optical-vortex scattering, which alters both helicity and OAM. Furthermore, we theoretically show that the relative efficiencies of these two scattering channels reverse as the degree of light focusing increases. This observed coexistence and reversal stand in stark contrast to the established paradigm of single-magnon-induced BLS in collimated or paraxial light~\cite{L2000,SN2000}. 

Schematic diagrams of the entire experimental setup are shown in Figs.~\ref{fig:setup111}(a) and \ref{fig:setup111}(b). 
A $0.5\text{-}\text{mm}$-diameter yttrium iron garnet (YIG) sphere was positioned at the center of the magnetic-circuit gap. It reached magnetic saturation upon the application of an external magnetic field of approximately $150\,\text{kA/m}$ along the $\langle111\rangle$ crystal axis instead of the previous $\langle100\rangle$ axis~\cite{RA2025}, and parallel to the $z$-axis. 
In this geometry, the cubic YIG crystal exhibits threefold rotational symmetry about the $z$-axis.
A single-loop coil, located directly above the YIG sphere, generated an oscillating magnetic field perpendicular to the saturation magnetization, thereby coupling the applied microwave field with the uniform magnon mode, conventionally known as the Kittel mode~\cite{C1948,LR1957}. 
Figure~\ref{fig:setup111}(d) shows the microwave reflection spectrum $|\text{S}_{11}|$ obtained using a vector network analyzer, which clearly indicates ferromagnetic resonance of the uniform mode. 
The resonance frequency of this mode was determined to be $\omega_m/2\pi = 4.10\,\text{GHz}$, and the number of excited magnons was estimated via a Lorentzian fit~\cite{RA2016}.

The experimental procedure was conducted as follows. 
First, the uniform-mode magnons within the YIG sphere were coherently excited via the single-loop coil.
A right-circularly polarized Gaussian beam, featuring a diameter of $50\,\mu\text{m}$ at the incident spherical surface and a wavelength of $1.55\,\mu\text{m}$, was directed into the sphere. 
The beam was gently focused using a convex lens with a focal length of $100\,\text{mm}$ positioned in front of the sphere, allowing the incident light to be treated as quasi-paraxial. 
The spherical surfaces at the incident and exit planes both functioned as convex lenses, each with a focal length of approximately $230\,\mu\text{m}$. 
Consequently, the input light formed a beam waist near the center of the sphere, closely coinciding with the geometric center. 
Within the sphere, strong focusing and subsequent divergence caused the light to propagate as a non-paraxial beam. 
Next, the light scattered by the uniform-mode magnons was recollimated into paraxial light using a multiple-lens system. 
Outside the sphere, the optical spin angular momentum (SAM, i.e., helicity) and OAM were analyzed. 
Finally, the specific scattering processes (Stokes or anti-Stokes) were determined, and the scattering efficiency of each process was quantitatively evaluated. 
Further methodological details are provided in the Appendix in our previous work~\cite{RA2025}.

The Laguerre-Gaussian $({\text{LG}}_{p}^{l_p})$ modes serve as a convenient set of basis vectors to describe the optical fields~\cite{LM1992}, where the index $l_p$ denotes the winding number and $(p+1)$ represents the number of radial nodes. 
In this study, we consider exclusively the case where $p=0$. 
Because the azimuthal phase term ($\exp(i l_p \phi)$) of the $\text{LG}$ mode gives rise to a helical wavefront, such a beam is referred to as an optical vortex. 
The winding direction of the helical wavefront is determined by the sign of the index $l_p$, and each photon in the $\text{LG}$ mode carries an OAM of $l_p \hbar$. 
The SAM per photon is given by $s_p \hbar$, where $s_p = \pm 1$ corresponds to left- or right-handed circularly polarized light, respectively. 
Since the $\text{LG}$ mode can simultaneously carry both SAM and OAM, we adopt the notation established in our pioneering work on magnon-induced optical vortex scattering~\cite{RA2025}: the optical mode is designated as ${\text{LG}}_{l_p,s_p}$, where the first and second subscripts represent the OAM and SAM quantum numbers, respectively.

\begin{figure}[t]
\begin{center}
\includegraphics[width=8.0cm,angle=0]{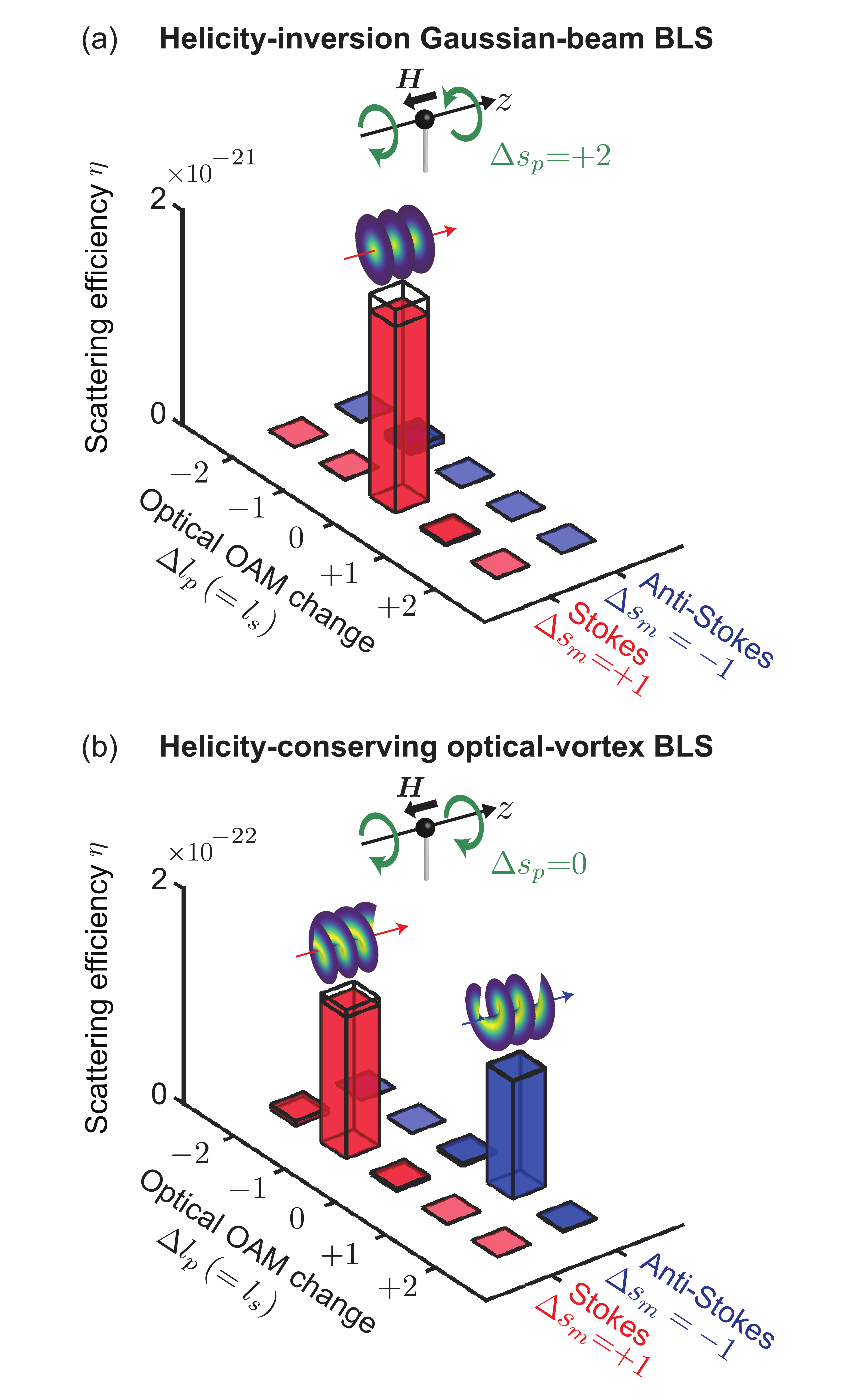}
\caption{
Scattering efficiencies of the Stokes and anti-Stokes sidebands.
Quantitative comparisons are shown for (a)~helicity-inversion Gaussian-beam Brillouin light scattering (BLS) and (b)~helicity-conserving optical-vortex BLS under an external magnetic field of $\bm{H} \parallel \langle111\rangle$. The red and blue bars represent the scattering efficiencies of the Stokes and anti-Stokes sidebands, respectively. 
The height of each colored bar indicates the mean scattering efficiency, while the difference between the top of the black wireframe and the bar denotes the standard deviation estimated from six independent measurements. 
Note that the vertical axes in (a) and (b) differ by one order of magnitude. 
The sign of each angular momentum quantum number is defined with respect to the quantization axis oriented in the positive $z$-direction.
}
\label{fig:eff_111}
\end{center}
\end{figure}

Figures~\ref{fig:eff_111}(a) and \ref{fig:eff_111}(b) show the measured magnon-induced BLS efficiencies obtained with an incident right-handed circularly polarized Gaussian beam. 
The scattering efficiency $\eta$ is defined as the probability that a single incident photon is scattered by a single magnon; it is experimentally derived from the optical beat signal at the angular frequency $\omega_R = \omega_m + \omega_A$ for Stokes scattering, and from the signal at $\omega_B = \omega_m - \omega_A$ for anti-Stokes scattering. 
Specific details regarding the calibration methodology are provided in Refs.~\cite{RA2016,RA2019}.
The sign and magnitude of the SAM $s_p$ per incident or scattered photon and the SAM $s_m$ per magnon are uniquely determined by their physical definitions and the experimental geometry. 
The OAM $l_s$ of the scattered photons was analyzed using our established method that leverages the spatial phase modulation characteristics of the spatial light modulator (SLM) and the spatial filtering of the SM fiber, as conceptually illustrated in Fig.~\ref{fig:setup111}(a) (see also Ref.~\cite{RA2025}). 
The changes in angular momentum for each scattering event are denoted as $\Delta s_p$, $\Delta s_m$, and $\Delta l_p$, respectively. Since the incident Gaussian beam carries no OAM, the OAM change $\Delta l_p$ is identical to the scattered OAM $l_s$. Consequently, $\Delta l_p = 0$ indicates a Gaussian-beam scattering process, whereas a finite $\Delta l_p$ signifies an optical-vortex scattering process. 

In Fig.~\ref{fig:eff_111}(a), a single significant Stokes sideband (red bar) with $\Delta l_p = 0$ is observed under the right-to-left circular polarization configuration, yielding a scattering efficiency of $1.7 \times 10^{-21}$. 
This corresponds to the helicity-inversion Gaussian-beam scattering process. 
Conversely, in Fig.~\ref{fig:eff_111}(b), under the helicity-conserving configuration, a prominent Stokes sideband with $\Delta l_p = -1$ (red bar) emerges with a scattering efficiency of $1.3 \times 10^{-22}$. 
This process represents the helicity-conserving optical-vortex scattering. Furthermore, another prominent anti-Stokes sideband (blue bar) with $\Delta l_p = +1$ simultaneously appears with a scattering efficiency of $1.1 \times 10^{-22}$. 
Taken together, these experimental results clearly demonstrate the coexistence of multiple distinct scattering pathways induced by uniform-mode magnons.

\begin{figure}[t]
\begin{center}
\includegraphics[width=8.0cm,angle=0]{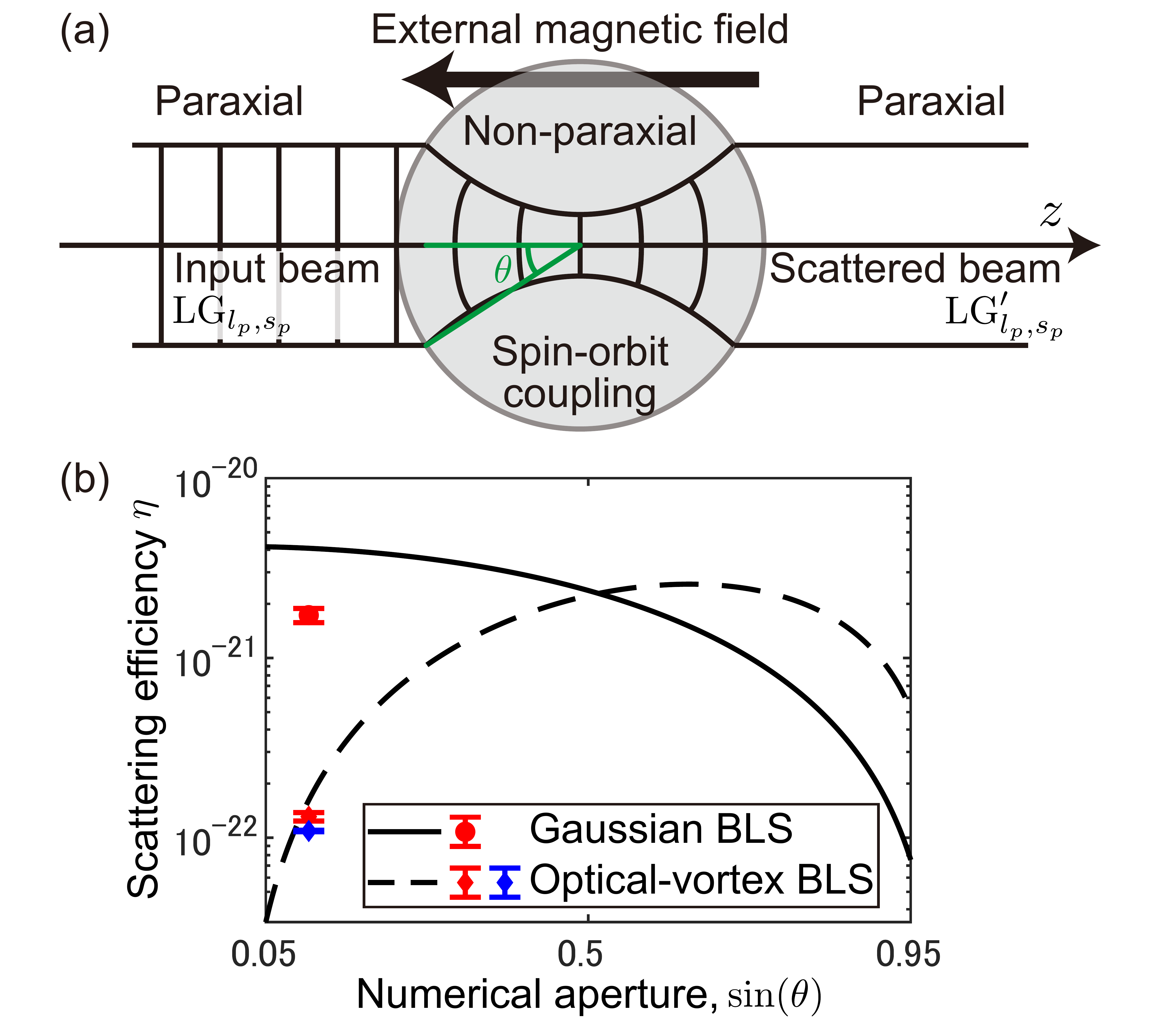}
\caption{
Normalized numerical aperture (NA) dependence of the scattering efficiencies. (a)~Schematic of light propagation around the spherical YIG sample. 
(b)~Theoretical curves and experimental data for the NA dependences of the helicity-inversion Gaussian-beam BLS (solid line), described by Eq.~(\ref{eq:eff_0m1_0p1}), and the helicity-conserving optical-vortex BLS (dashed line), described by Eq.~(\ref{eq:eff_0p1_m1m1}). The experimentally observed scattering efficiencies with their respective standard deviations [from Figs.~\ref{fig:eff_111}(a) and \ref{fig:eff_111}(b)] are plotted at $\sin\theta = 0.11$. 
Red and blue denote the Stokes and anti-Stokes processes, respectively.
This experimental NA value is estimated from the incident beam diameter of $50\,\mu\text{m}$ and the effective focal length of $230\,\mu\text{m}$.
}
\label{fig:calc}
\end{center}
\end{figure}

To discuss the transfer of angular momentum in the observed scattering processes, we adopt a quantum mechanical perspective and focus on the elementary processes of magnon-induced BLS~\cite{T1968,PR1968}. 
Because these elementary processes can be modeled as three-wave mixing events involving one magnon and two photons, the change in total angular momentum during BLS is expressed as $\Delta s_m + \Delta s_p + \Delta l_p$.
It can be readily verified that the helicity-conserving vortex scattering processes shown in Fig.~\ref{fig:eff_111}(b) satisfy the conventional conservation of total angular momentum: $\Delta s_m + \Delta s_p + \Delta l_p = 0$. 
In contrast, for the helicity-inversion Gaussian-beam scattering depicted in Fig.~\ref{fig:eff_111}(a), this conventional conservation law appears to be violated. 
However, previous research has demonstrated that in angular momentum transfer processes between ferromagnetic magnons and photons within crystals lacking continuous rotational symmetry, the crystal angular momentum can play a decisive role~\cite{RA2019}. 
Since this situation is physically analogous to cases where the reciprocal lattice vector intervenes in linear-momentum conservation laws for electrons or phonons, it is referred to as a rotational Umklapp process. 
In our specific system involving a cubic crystal under the Faraday geometry ($\bm{H} \parallel \langle111\rangle$), the crystal angular momentum is restricted to integer multiples of $3\hbar$. 
Taking this discrete symmetry into account, the conservation of total angular momentum must be generalized to the following selection rule: 
\begin{equation} \Delta s_m + \Delta s_p + \Delta l_p \equiv 0 \pmod 3, \end{equation} which successfully accounts for and confirms the conservation in the helicity-inversion process shown in Fig.~\ref{fig:eff_111}(a).
To validate this mechanism, we modeled the tightly focused optical system. Upon refraction at the YIG spherical interface, the incident paraxial $\text{LG}_{l_p,s_p}$ mode converts into a linear combination of spin-orbit-coupled (SO) modes conserving the total angular momentum $l_p+s_p$~\cite{KE2011}:
\begin{align}
{\text{LG}}_{l_p,s_p} = \sqrt{\cos \theta} \Big[ a \times {\text{LG}}_{l_p,s_p}^{\text{SO}} - b \times {\text{LG}}_{l_p+2s_p,-s_p}^{\text{SO}} \notag \\  - \sqrt{2ab} \times {\text{LG}}_{l_p+s_p,0}^{\text{SO}} \Big], \label{eq:OSOI}
\end{align}
where $a = \cos^2(\theta/2)$, $b = \sin^2(\theta/2)$, and $\theta$ is the half-aperture angle [Fig.~\ref{fig:calc}(a)]. The third term introduces a longitudinal electric field component that grows with $\theta$ but reverts to paraxial modes upon exiting the sphere. 

Our framework attributes the helicity-conserving scattering [Fig.~\ref{fig:eff_111}(b)] to transitions between these SO modes driven by uniform-mode magnons. As detailed in Appendix~\ref{sec:shcop}, the Stokes channel (${\text{LG}}_{0,-1} \rightarrow {\text{LG}}'_{-1,-1}$, where the prime denotes the scattered mode) arises from the magneto-optical transition from ${\text{LG}}_{-1,0}^{\text{SO}}$ to ${\text{LG}}'^{\text{SO}}_{-1,-1}$, yielding a calculated efficiency of $1.6 \times 10^{-22}$, identical to that of the anti-Stokes channel. Conversely, the Stokes helicity-inversion Gaussian scattering (${\text{LG}}_{0,-1} \rightarrow {\text{LG}}'_{0,+1}$) in Fig.~\ref{fig:eff_111}(a) requires the rotational Umklapp process. 
Appendix~\ref{sec:shig} details this transition from ${\text{LG}}_{0,-1}^{\text{SO}}$ to ${\text{LG}}'^{\text{SO}}_{0,+1}$, calculating an efficiency of $4.1 \times 10^{-21}$. As shown in Fig.~\ref{fig:calc}(b), both theoretical calculations are in good agreement with the experimental data, confirming the interplay of magneto-optics, optical spin-orbit interaction, and rotational Umklapp processes.

To further elucidate the distinct physical characteristics of the coexisting Gaussian-beam and optical-vortex BLS, we turn our attention to the normalized numerical aperture ($\text{NA}$) dependence of the scattering efficiencies.
Here, we adopt the standard optical definition $\text{NA} = \sin\theta$~\cite{1999ME}, which scales monotonically with the focus angle. 
Figure~\ref{fig:calc}(b) plots the calculated NA dependence alongside the experimentally measured efficiencies. 
Under our specific experimental condition ($\text{NA} \approx 0.11$), the Gaussian-beam BLS exhibits a higher scattering efficiency than the optical-vortex BLS. However, the theoretical curves in Eqs.~(\ref{eq:eff_0p1_m1m1}) and (\ref{eq:eff_0m1_0p1}) predict that this relationship reverses at a critical $\text{NA}$ of approximately $0.5$.
The physics unveiled here is highly general; it should hold true not only for the present configuration of paraxial light incident on a ferromagnetic sphere, but also for configurations where focused light is incident on ferromagnetic thin films or nanostructures. 
For instance, it has only recently been recognized that treating light as simple parallel or plane waves is fundamentally inadequate for describing BLS under tight-focusing conditions~\cite{KM2026}.
Consequently, it is crucial to account for not only the polarization state (spin) of the optical field but also its vortex state, i.e., the spatial phase and amplitude distributions of the complex electric field.

Finally, we emphasize that the generalized conservation law in Eq.~(1) is a necessary condition, but by no means sufficient. 
For instance, the quantum-number combination $(\Delta s_m, \Delta s_p, \Delta l_p) = (-1, +2, +2)$ satisfies Eq.~(1) but is absent in Fig.~2(a). 
Our microscopic theory reveals that its matrix element vanishes identically, strictly prohibiting this transition even with the rotational Umklapp process. 
Conversely, the combinations $(\pm1, 0, \pm2)$ also satisfy Eq.~(1) but remain unobserved in Fig.~2(b); as detailed in Appendix~\ref{sec:sahcov}, our calculations show that their scattering efficiencies are identical at $3.8 \times 10^{-26}$, demonstrating that this absence is simply due to their inherently low efficiency.

In summary, by utilizing ferromagnetic spheres, we have demonstrated the coexistence of two distinct scattering pathways: rotational-Umklapp-process-assisted Gaussian-beam scattering accompanied by helicity inversion, and optical-vortex scattering without helicity inversion. In particular, the microscopic mechanism of the former process has been quantitatively elucidated by simultaneously incorporating the rotational Umklapp process, magneto-optical effects, and optical spin-orbit interactions. Furthermore, we have revealed that the dominance between these two coexisting scattering processes critically depends on the degree of light focusing.

\begin{acknowledgments}
We are grateful to Y.~Nakamura for providing the YIG sphere used in this study and to A.~Osada for fruitful discussions. 
This work was supported by the Japan Science and Technology Agency (JST) PRESTO (Grant No.~JPMJPR200A to R.H.); 
JST ASPIRE (Grant No.~JPMJAP2409 to T.O.); 
the Japan Society for the Promotion of Science (JSPS) KAKENHI (Grant Nos.~JP22K14589 to R.H., JP24H0223A to Y.S., and JP25K00937 to R.H.); and the Cooperative Research Project of the Research Institute of Electrical Communication (RIEC), Tohoku University (to T.O.). 
The authors declare no competing financial interests.
\end{acknowledgments}

\vspace{1em}
\noindent\textbf{Data Availability} --- The data that support the findings of this study are available from the corresponding author upon reasonable request.

\clearpage 
\widetext 

\appendix 

\setcounter{equation}{0} 
\setcounter{figure}{0} 
\setcounter{table}{0} 
\renewcommand{\theequation}{\thesection\arabic{equation}} 
\renewcommand{\thefigure}{\thesection\arabic{figure}}     
\renewcommand{\thetable}{\thesection\arabic{table}}       

\makeatletter 
\@ifundefined{theHequation}{}{\renewcommand{\theHequation}{Appendix.\theequation}} 
\@ifundefined{theHfigure}{}{\renewcommand{\theHfigure}{Appendix.\thefigure}} 
\@ifundefined{theHtable}{}{\renewcommand{\theHtable}{Appendix.\thetable}} 
\makeatother

\section{Scattering efficiency for Stokes helicity-conserving optical-vortex scattering (${\text{LG}}_{0,-1}\rightarrow{\text{LG}'}_{-1,-1}$)} 
\label{sec:shcop}
In this section, we derive the scattering efficiency for the Stokes helicity-conserving optical-vortex scattering process (${\text{LG}}_{0,-1}\rightarrow{\text{LG}}'_{-1,-1}$). Using Eq.~(2) in the main text, we establish the correspondence between the far-field paraxial modes and the spin-orbit-coupled (SO) light modes inside the ferromagnetic sphere. For the incident Gaussian paraxial mode and the scattered paraxial vortex mode, the fields are expressed as
\begin{equation}
{\text{LG}}_{0,-1} = \sqrt{\cos\theta} \left( a \times {\text{LG}}_{0,-1}^{\text{SO}} - b \times {\text{LG}}_{-2,+1}^{\text{SO}} - \sqrt{2ab} \times {\text{LG}}_{-1,0}^{\text{SO}} \right), \label{eq:LG0m1_OptSOI}
\end{equation}
and
\begin{equation}
{\text{LG}'}_{-1,-1} = \sqrt{\cos\theta} \left( a \times {\text{LG}'}_{-1,-1}^{\text{SO}} - b \times {\text{LG}'}_{-3,+1}^{\text{SO}} - \sqrt{2ab} \times {\text{LG}'}_{-2,0}^{\text{SO}} \right), \label{eq:LGm1m1_OptSOI}
\end{equation}
respectively. Following the microscopic framework outlined in Ref.~\cite{RA2025}, the Stokes scattering channel is mediated by the transition from the ${\text{LG}}_{-1,0}^{\text{SO}}$ mode to the ${\text{LG}}'^{\text{SO}}_{-1,-1}$ mode, which share the same OAM quantum number. By evaluating the overlapping matrix elements, the total scattering efficiency $\eta$ is given by
\begin{equation}
\eta = 2 a^3 b \cos^2\theta \, \left[ \frac{\Omega_0 \Omega' l^2 \mu_B^2 n}{128 \pi V_s c^2} \left( f + \frac{\mu_B N}{3 V_s} (G_{11} - G_{12} + G_{44}) \right)^2 \right],
\label{eq:eff_0p1_m1m1}
\end{equation}
where the parameters are defined in Table~\ref{table:known_values_111}. Utilizing these literature values, the scattering efficiency is quantitatively evaluated to be $1.6 \times 10^{-22}$.

\section{Scattering efficiency for Stokes helicity-inversion Gaussian-beam scattering (${\text{LG}}_{0,-1}\rightarrow{\text{LG}}'_{0,+1}$)} 
\label{sec:shig}

In this section, we derive the scattering efficiency for the Stokes helicity-inversion Gaussian-beam scattering process (${\text{LG}}_{0,-1}\rightarrow{\text{LG}}'_{0,+1}$). Using Eq.~(2) in the main text, the correspondence between the far-field paraxial mode and the spin-orbit-coupled (SO) light modes inside the ferromagnetic sphere for the scattered paraxial mode is expressed as
\begin{equation}
{\text{LG}'}_{0,+1} = \sqrt{\cos\theta} \left( a \times {\text{LG}'}_{0,+1}^{\text{SO}} - b \times {\text{LG}'}_{+2,-1}^{\text{SO}} - \sqrt{2ab} \times {\text{LG}'}_{+1,0}^{\text{SO}} \right). \label{eq:LGm3p1_OptSOI}
\end{equation}
According to the microscopic framework outlined in Ref.~\cite{RA2025}, this Stokes scattering channel is mediated by the transition from the incident ${\text{LG}}_{0,-1}^{\text{SO}}$ mode [Eq.~(\ref{eq:LG0m1_OptSOI})] to the scattered ${\text{LG}}'^{\text{SO}}_{0,+1}$ mode. Importantly, since these two modes have optical spin angular momentum (SAM) quantum numbers that differ by exactly 2, this transition cannot occur via conventional magneto-optical coupling; instead, it is explicitly enabled by the rotational Umklapp process arising from the discrete crystal symmetry. By evaluating the overlapping matrix elements via this lattice-driven angular momentum transition, the total scattering efficiency $\eta$ is given by
\begin{equation}
\eta = a^4 \cos^2\theta \, \left[ \frac{\Omega_0 \Omega' l^2 \mu_B^2 n}{128 \pi V_s c^2} \left( \frac{\mu_B N}{3 V_s} (G_{11} - G_{12} - 2G_{44}) \right)^2 \right],
\label{eq:eff_0m1_0p1}
\end{equation}
where the parameters are defined in Table~\ref{table:known_values_111}. Utilizing these literature values, the scattering efficiency of this Umklapp-assisted process is quantitatively evaluated to be $4.1 \times 10^{-21}$.

\section{Scattering efficiency for Stokes or anti-Stokes helicity-conserving optical-vortex scattering ($\mathrm{LG}_{0,-1}\rightarrow\mathrm{LG}'_{+2,-1}$ or $\mathrm{LG}'_{-2,-1}$)} 
\label{sec:sahcov}
In this section, we evaluate the scattering efficiency for the Stokes and anti-Stokes helicity-conserving optical-vortex scattering processes ($\mathrm{LG}_{0,-1}\rightarrow\mathrm{LG}'_{+2,-1}$ or $\mathrm{LG}'_{-2,-1}$). 
By utilizing Eq.~(2) in the main text, the correspondence between the far-field paraxial mode and the spin-orbit-coupled (SO) light modes inside the ferromagnetic sphere for the scattered paraxial mode can be analytically expressed as
\begin{equation}
\mathrm{LG}'_{+2,-1} = \sqrt{\cos\theta} \left( a \times \mathrm{LG}'^{\mathrm{SO}}_{+2,-1} - b \times \mathrm{LG}'^{\mathrm{SO}}_{0,+1} - \sqrt{2ab} \times \mathrm{LG}'^{\mathrm{SO}}_{+1,0} \right). \label{eq:LGp2m1_OptSOI}
\end{equation}
According to the microscopic framework outlined in Ref.~\cite{RA2025}, this Stokes scattering channel is mediated by the transition from the incident $\mathrm{LG}_{0,-1}^{\mathrm{SO}}$ mode [Eq.~(\ref{eq:LG0m1_OptSOI})] to the scattered $\mathrm{LG}'^{\mathrm{SO}}_{0,+1}$ mode. 
Importantly, since these two internal modes possess optical spin angular momentum (SAM) quantum numbers that differ by exactly 2, this transition is strictly forbidden via conventional magneto-optical coupling; instead, it is explicitly enabled by the rotational Umklapp process arising from the discrete crystal symmetry. By evaluating the overlapping matrix elements for this lattice-driven angular momentum transition, the total scattering efficiency $\eta$ is derived as
\begin{equation}
\eta = a^2 b^2 \cos^2\theta \, \left[ \frac{\Omega_0 \Omega' l^2 \mu_B^2 n}{128 \pi V_s c^2} \left( \frac{\mu_B N}{3 V_s} (G_{11} - G_{12} - 2G_{44}) \right)^2 \right],
\label{eq:eff_0m1_0p1_another}
\end{equation}
where the parameters are defined in Table~\ref{table:known_values_111}. Utilizing these literature values, the scattering efficiency of this Umklapp-assisted process is quantitatively calculated to be $3.8 \times 10^{-26}$.
We note that the scattering efficiency for the corresponding anti-Stokes process yields an identical expression and value to that given in Eq.~(\ref{eq:eff_0m1_0p1_another}).

\begin{table}[t]
  \centering
  \caption{Literature and known physical parameters used in the theoretical calculations.} 
  \label{table:known_values_111}
  \begin{tabular}{lc} 
    \hline\hline 
    Physical quantity & Value \\ 
    \hline 
    Aperture angle $\theta$ & $0.11\,\text{rad}$ \\ 
    Angular frequency of incident light $\Omega_0$ & $2\pi\times193\,\text{THz}$ \\ 
    Angular frequency of scattered light $\Omega'$ & $2\pi\times(193 \pm 0.0041)\,\text{THz}$ \\ 
    Interaction length $l$ & $0.5\,\text{mm}$ \\ 
    Spin density of YIG $n$~\cite{DA2009} & $2.1 \times 10^{28}\,\text{m}^{-3}$ \\ 
    Sample volume $V_s$ & $\displaystyle \frac{4}{3}\pi\left(\frac{l}{2}\right)^3$ \\ 
    Faraday coefficient of YIG $f$~\cite{D1991} & $\displaystyle \frac{2\sqrt{\epsilon_r}\nu}{k_0(-\frac{1}{2}\mu_B n)}$ \\ 
    Relative permittivity of YIG $\epsilon_r$~\cite{DA2009} & $2.2$ \\ 
    Verdet constant of YIG $\nu$~\cite{MS2000} & $380\,\text{rad/m}$ \\ 
    Cotton-Mouton coefficient $G_{44}\left(\frac{\mu_B}{2}n\right)^2$~\cite{DA2009} & $-1.14 \times 10^{-4}$ \\ 
    Cotton-Mouton coefficient $\Delta g\left(\frac{\mu_B}{2}n\right)^2 = \left(G_{11}-G_{12}-2G_{44}\right)\left(\frac{\mu_B}{2}n\right)^2$~\cite{DA2009} & $5.73 \times 10^{-5}$ \\ 
    \hline\hline
  \end{tabular}
\end{table}







\end{document}